\documentstyle[epsf,astrcite]{mn}

\newif\ifAMStwofonts

\newcommand{\un}[2]{\mbox{\rm\thinspace #1$^{#2}$}}

\newcommand{\be}[1]{\begin{equation}\label{#1}}
\newcommand{\ee}{\end{equation}}

\newcommand{\degs}{\mbox{$^{\circ}$}}
\newcommand{\Eq}[1]{Eq.\,(\ref{#1})}

\newcommand{\Fig}[1]{Fig.\,\ref{#1}}

\newcommand{\gsim}{\mathrel{\hbox{\rlap{\lower.55ex \hbox {$\sim$}}
                   \kern-.3em \raise.4ex \hbox{$>$}}}}
\newcommand{\lsim}{\mathrel{\hbox{\rlap{\lower.55ex \hbox {$\sim$}}
                   \kern-.3em \raise.4ex \hbox{$<$}}}}

\newcommand{\Sect}[1]{Sect.\,\ref{#1}}

\newcommand{\spr}[1]{^{\rm #1}}
\newcommand{\sub}[1]{_{\rm #1}}

\newlength{\ffh}

\newcommand{\photcms}{\un{phot}{}\un{cm}{-2}\un{s}{-1}}

\ifoldfss
  \ifCUPmtlplainloaded \else
    \NewTextAlphabet{textbfit} {cmbxti10} {}
    \NewTextAlphabet{textbfss} {cmssbx10} {}
    \NewMathAlphabet{mathbfit} {cmbxti10} {} 
    \NewMathAlphabet{mathbfss} {cmssbx10} {} 
  \fi
  \ifAMStwofonts
    \ifCUPmtlplainloaded \else
      \NewSymbolFont{upmath} {eurm10}
      \NewSymbolFont{AMSa} {msam10}
      \NewMathSymbol{\upi}     {0}{upmath}{19}
      \NewMathSymbol{\umu}     {0}{upmath}{16}
      \NewMathSymbol{\upartial}{0}{upmath}{40}
      \NewMathSymbol{\leqslant}{3}{AMSa}{36}
      \NewMathSymbol{\geqslant}{3}{AMSa}{3E}

       \let\ge=\geqslant
    \fi
  \fi
\fi 

\ifnfssone
  \newmathalphabet{\mathit}
  \addtoversion{normal}{\mathit}{cmr}{m}{it}
  \addtoversion{bold}{\mathit}{cmr}{bx}{it}
  \newmathalphabet{\mathbfit} 
  \addtoversion{normal}{\mathbfit}{cmr}{bx}{it}
  \addtoversion{bold}{\mathbfit}{cmr}{bx}{it}
  \newmathalphabet{\mathbfss} 
  \addtoversion{normal}{\mathbfss}{cmss}{bx}{n}
  \addtoversion{bold}{\mathbfss}{cmss}{bx}{n}
  \ifAMStwofonts
    \ifCUPmtlplainloaded \else
      \UseAMStwoboldmath
      \makeatletter
      \new@mathgroup\upmath@group
      \define@mathgroup\mv@normal\upmath@group{eur}{m}{n}
      \define@mathgroup\mv@bold\upmath@group{eur}{b}{n}
      \edef\UPM{\hexnumber\upmath@group}
      \new@mathgroup\amsa@group
      \define@mathgroup\mv@normal\amsa@group{msa}{m}{n}
      \define@mathgroup\mv@bold\amsa@group{msa}{m}{n}
      \edef\AMSa{\hexnumber\amsa@group}
      \makeatother
      \mathchardef\upi="0\UPM19
      \mathchardef\umu="0\UPM16
      \mathchardef\upartial="0\UPM40
      \mathchardef\leqslant="3\AMSa36
      \mathchardef\geqslant="3\AMSa3E

       \let\ge=\geqslant
    \fi
  \fi
\fi 

\ifnfsstwo
  \DeclareMathAlphabet{\mathbfit}{OT1}{cmr}{bx}{it}
  \SetMathAlphabet\mathbfit{bold}{OT1}{cmr}{bx}{it}
  \DeclareMathAlphabet{\mathbfss}{OT1}{cmss}{bx}{n}
  \SetMathAlphabet\mathbfss{bold}{OT1}{cmss}{bx}{n}
  \ifAMStwofonts
    \ifCUPmtlplainloaded \else
      \DeclareSymbolFont{UPM}{U}{eur}{m}{n}
      \SetSymbolFont{UPM}{bold}{U}{eur}{b}{n}
      \DeclareSymbolFont{AMSa}{U}{msa}{m}{n}
      \DeclareMathSymbol{\upi}{0}{UPM}{"19}
      \DeclareMathSymbol{\umu}{0}{UPM}{"16}
      \DeclareMathSymbol{\upartial}{0}{UPM}{"40}
      \DeclareMathSymbol{\leqslant}{3}{AMSa}{"36}
      \DeclareMathSymbol{\geqslant}{3}{AMSa}{"3E}

       \let\ge=\geqslant
    \fi
  \fi
\fi 

\ifCUPmtlplainloaded \else
  \ifAMStwofonts \else 
    \def\upi{\pi}
    \def\umu{\mu}
    \def\upartial{\partial}
  \fi
\fi

\title[Detecting M31 gamma-ray bursts]{Detecting gamma-ray bursts from
       M31 with the wide field X-ray cameras on board BeppoSAX}
\author[M. Ruszkowski and R. A. M. J. Wijers]
       {Mateusz Ruszkowski$^{1,2,3}$ and Ralph A. M. J. Wijers$^1$ \\
        $^1$Institute of Astronomy, Madingley Road, Cambridge, CB3 0HA\\
        $^2$Warsaw University Astronomical Observatory, Al.\ Ujazdowskie 4,
	00-478 Warsaw, Poland\\
	$^3$Nicolaus Copernicus Astronomical Center, Polish Academy of Sciences,
	    Bartycka 18, 00-716 Warsaw, Poland\\
        Email: {\tt ruszkows@sirius.astrouw.edu.pl}  and 
	       {\tt ramjw@ast.cam.ac.uk}\\}
\date{Submitted to MNRAS, 21 January 1997}
\pubyear{1996}

\begin{document}

\maketitle

\newcommand{\fmmm}[1]{\mbox{$#1$}}
\newcommand{\scnd}{\mbox{\fmmm{''}\hskip-0.3em .}}
\newcommand{\scnp}{\mbox{\fmmm{''}}}

\begin{abstract}
Gamma-ray bursters emit a small fraction of their flux in X rays, and because
X-ray detectors are often very sensitive they may probe the gamma-ray burst
universe more deeply than the current best gamma-ray instruments.  On the
reasonable assumptions that spectra of bursts observed by BATSE may be used
to predict the X-ray fluxes of gamma-ray bursts, and that any corona of
bursts around M31 is similar to the one around the Milky Way, we predict
the rate at which the wide field cameras on board BeppoSAX should detect
bursts from the Milky Way and M31. These rates are such that a one-month
observation of M31 would have to either detect bursts from M31 or exclude
most galactic models of gamma-ray bursts. (It is shown how the remainder
can be dealt with.) Therefore such an
observation would settle the long-standing dispute over their location.
\end{abstract}


   \section{Introduction}
   \label{intro}

The results of the BATSE mission (see Fishman and Meegan
1995)\nocite{fm:95}, combined with earlier data sets for bright bursts such
as the one collected by PVO (Fenimore et~al.\ 1993)\nocite{fehkl:93} have
shown that (1) gamma-ray burst positions are distributed uniformly and
randomly on the sky (Briggs et~al.\ 1996a,b)\nocite{bppmf:96,bppmf2:96} 
and (2)  the cumulative number as a function of peak
flux, $N(>P_\gamma)$, is consistent with a constant rate density of bursts
within some volume around us, and a decreasing density outside that volume.
This implies that we are at the centre of a gamma-ray burst universe of
which we can see the edge and which looks the same in all directions. Most
distance scales are therefore excluded.  The first remaining one is the
high-redshift universe, with the edge being caused either by cosmological
volume effects near and beyond $z=1$ or by evolution of the density at
moderate redshift (or both).  The second one is an extended corona of our
Galaxy, much bigger than the dark-matter halo and invented for the purpose
of housing gamma-ray bursts (GRB). We are not strictly in its centre, but the
average GRB distance can be made large enough that the anisotropy due to
our offset from the centre is below the limit set by the BATSE data on
burst positions. At the
same time it can still be small enough that we need not see M31
(Briggs et~al.\ 1996b). The aim of
this paper is to demonstrate the capability of the Wide Field Cameras (WFC)
on board BeppoSAX (launched in April 1996) to distinguish between these
options by searching for the hypothetical corona of GRB around M31.
We first discuss the Z-ray detectability of GRB (Sect.~\ref{xgrb}) and our
implementation of corona models (Sect.~\ref{coro}). Our results are
presented in Sect.~\ref{resu} and compared to previous results in
Sect.~\ref{disc}.

   \section{X-ray detection of GRB}
   \label{xgrb}

Ginga observations and some earlier detections indicate that gamma-ray
bursts emit some X rays (for an overview of early X-ray detections, see
Preece et~al 1996)\nocite{pbppm:96}.  It is only a small fraction of the
flux (2\% or so median; Laros et~al.\ 1984, Yoshida et~al.\
1989\nocite{lefks:84,ymint:89}), but since X- and gamma-ray instruments
are photon counting devices it is the higher count rate ratio of X rays to
gamma rays that matters. In addition, X-ray detectors usually have lower
backgrounds because of their imaging capability. In all, they can see some
fainter bursts than BATSE, the currently most sensitive gamma-ray detector
looking for GRB, at a price of having a much smaller field of view.  This
was recently used by Hamilton et~al.\ (1996)\nocite{hgh:96} to constrain
galactic-coronal models of GRB using archival Einstein data. They used
galaxies typically a few Mpc away, the compromise distance for Einstein's
exquisite sensitivity but very small field of view.  The lower sensitivity
of the WFC mean we should observe more nearby galaxies to constrain 
galactic-coronal models with them; the large field of view means that we
can go as close as M31 without losing too many bursts because they lie
outside the field of view.

      \subsection{The WFCs on board SAX}
      \label{xgrb.wfc}

There are two WFCs on board SAX, looking at opposite directions on the sky
and perpendicular to the on-axis instruments. They are coded-mask imaging
instruments with an entry mask of 256$\times$256 1\,mm$^2$ pixels placed
700\,mm away from the detector plane. This leads to a response that is
approximately
triangular in both $x$ and $y$ and falls to zero 20$^\circ$ away from the
optical axis along the $x$ and $y$ directions. It is sensitive to the
energy range 1.6--32\,keV. The response function was kindly supplied to us
by Dr.~J. Heise of SRON Utrecht.  The angular resolution is a few
arcminutes for bright sources. The 130 cts/s background of the instrument
is mainly due to the diffuse extragalactic emission integrated over the
field of view. The instrumental backgrounds are small and stable due to
the low equatorial orbit which avoids the radiation belts and the South
Atlantic Anomaly.

      \subsection{The X-ray fluxes of GRB}
      \label{xgrb.xflu}

\begin{figure}
   \epsfxsize=\columnwidth\epsfbox{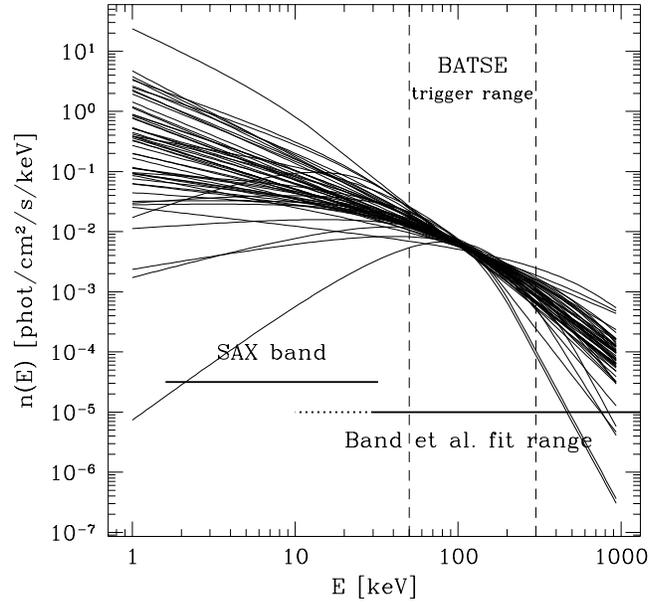}
   \caption{The best-fit model spectra for the GRB sample from Band et~al.
            with their extrapolation into the WFC band.  Note that the
            range over which the model fits were made significantly overlaps
            with the WFC band.
            \label{fi.grbspec}
            }
\end{figure}

\begin{figure}
   \epsfxsize=\columnwidth\epsfbox{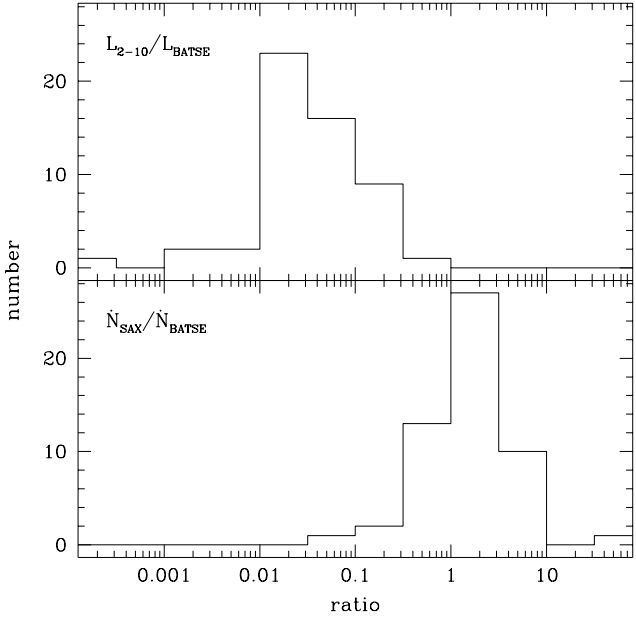}
   \caption{The distribution of classical X-ray flux (2--10\,keV) to BATSE
	    (50--300\,keV) flux ratio (top) and the WFC (1.6--32\,keV) to
	    BATSE photon number flux ratio, which is more relevant for
	    burst detection. The large difference is partly due to the
	    energy per photon being much less in X rays, and partly due to
	    the fact that the WFC band is much wider than the classical
	    X-ray band.
            \label{fi.ratios}
            }
\end{figure}

Previous workers have used a mean flux
ratio of typically 2\% between the BATSE flux and the X-ray flux of a
gamma-ray burst. Rather than rely on the few X-ray detected gamma-ray
bursts, we note that the WFC and BATSE sensitivity ranges overlap in the
10--30\,keV range, and that therefore extrapolating the BATSE spectra into
the WFC band should give reasonably good estimates of the expected SAX WFC
count rates. Band et~al.\ (1993) published detailed spectral fits to  a
sample of bright GRB from the first-year BATSE catalogue.
The model consists of two power laws connected by a smooth transition at a
break energy. Almost all break energies are well within the BATSE range.
The model fits are shown in \Fig{fi.grbspec}, scaled to a photon number
flux of 1\,phot\un{cm}{-2}\un{s}{-1}\ integrated over the 50--300\,keV
band.  (The integrated count rate over this band is used by BATSE to
trigger bursts.) We folded each of their 54 best-fit models through the WFC
response matrix and computed the ratio $\Phi\sub{WFC}$ of the WFC on-axis
count rate to the 50--300\,keV photon number flux. This may seem an odd
ratio to take, but the WFC detectability is determined by the count rate,
whereas the GRB flux distribution seen by BATSE is usually reported after
correction for detector response, i.e.\ as a photon number flux.  A
histogram of these ratios, and of the more commonly used
$F(2-10\un{keV}{})/F(50-300\un{keV}{})$, is shown in \Fig{fi.ratios}. The
median flux ratio agrees with the Ginga estimates of 2--6\%
(Yoshida et~al.\ 1989)\nocite{ymint:89}.  But the large spread is
crucial because it implies that a substantial fraction of GRB will be
brighter than previous authors have estimated using a constant
$F\sub{X}/F_\gamma$ and will therefore be more easily detected.

A possible concern is the extent to which our sample is representative of
the population. The selection criterion employed by Band et~al.\
(1993)\nocite{bmfsp:93} to obtain their subsample of the first BATSE
catalogue for spectral fitting is essentially a brightness cut, and the
authors consider the sample to be effectively a complete flux-limited one.
This reduces the question of fair sampling to whether the spectral shape
systematically changes with flux. There is one known effect of this kind,
namely that faint bursts have break energies that are smaller by a factor 2
than bright ones (Malozzi et~al.\ 1995)\nocite{mppbp:95}. Since our
spectral sample is from bright bursts, we may calculate the effect on the
values of $\Phi\sub{WFC}$ by decreasing the value of the break flux by a
factor 2 for the whole sample. Each value of $\Phi\sub{WFC}$ changes
differently because the change depends on the spectral slopes, but the
resulting distribution of $\Phi\sub{WFC}$ is shifted to higher values by a
factor 2, i.e.\ faint bursts are relatively brighter in X rays.

Moreover, if there is a correlation between spectral shape and luminosity,
and this is translated into a weak dependence of spectral shape on flux via
the selection bias on flux, we probably should not even want to correct it
because we will only see the most luminous bursts from M31 with SAX, so a
sample of spectra that favours luminous bursts is a better one to use.
Similarly, since the sample is gamma-ray selected, any bias in the spectral
sample will be towards GRB that are gamma-ray bright, so the true
population is likely to have on average greater X-ray brightnesses if the
sample is not fair.

In what follows we shall use the $\Phi\sub{WFC}$ distribution from the Band
et~al.\ sample to  estimate the X-ray detectability of GRB. We note that
all the known and potential biases discussed would increase the number of
detectable bursts from M31 over the calculations presented below.

      \subsection{X-ray detectability given a gamma-ray flux}
      \label{xgrb.xdet}

To decide whether a given GRB will be detected by the WFC, we assume that it is
a standard candle in the BATSE band. It has already been shown (Hakkila
et~al.\ 1995)\nocite{hmphb:95} that the range of gamma-ray luminosities of
GRB must be small for all corona models that are still viable. Ulmer and
Wijers (1995)\nocite{uw:95} also showed that for most luminosity functions,
the luminosity distribution of detected GRB is narrow even if the intrinsic
luminosity function of the population is not. If the luminosity function is
wide it increases the spread in X-ray luminosities at constant mean and
therefore helps X-ray detection, unless there is a strong correlation
between luminosity and spectral slope, so the standard-candle assumption
will yield the lowest estimate of the number of X-ray detectable GRB.

The procedure to decide whether a model GRB will be detected by the WFC is then
as follows. First, the 50--300\,keV peak flux $P_\gamma$ is given by the
model. Then the WFC on-axis count rate is computed by multiplying with
$\Phi\sub{WFC}$. The model also specifies the location of the burst, from
which we calculate the position $(x,y)$ of its image in the detector plane
(to be precise, the intersection of the line connecting the burst location
and the centre of the mask with the detector plane). The count rate due to
a burst at $(x,y)$ is less than that due to one on axis by a factor
$R(x,y)$, which accounts for the fact that it only illuminates a fraction
of the detector and for the usual factor $\cos\theta$ to account for the
fact that the detector plane is not perpendicular to the direction to the
burst:
\be{eq.rxy}
  R(x,y) = (1-\frac{|x|}{256})(1-\frac{|y|}{256})\cos\theta.
\ee
($x$ and $y$ are measured in mm from an origin at the detector centre, and
the $x$ and $y$ axes are parallel to the edges of the square mask.)
Given an integration time $T$ and background count rate $b$, we get
the total number of source counts $S=fP_\gamma\Phi\sub{WFC}RT$ and
background counts $B=bT$. The factor $f$ is required because $P_\gamma$ is
the peak flux, which will generally not be sustained for the full time $T$.
In terms of the instantaneous number flux $p_\gamma(t)$ from the
burst (assumed 0 outside the interval $(0,T)$) we can
formally define $f$ as
\be{eq.fdef}
  f\equiv\frac{\int_0^Tp_\gamma(t){\rm\, d}t}{P_\gamma T},
\ee
Obviously, for a fixed value of $T$ we will find a different $f$ for each
burst and $f$ will usually be smaller for shorter bursts.
We assume the background noise is Poissonian
and large enough to be approximated by a normal distribution, so we
can express the requirement that the burst be more than $\sigma$ standard
deviations above the background as a constraint on $P_\gamma$:
\be{eq.pmindef}
  P_\gamma \ge P\sub{\gamma}\spr{min}\equiv 
	\frac{\sigma\sqrt{b}}{f\Phi\sub{WFC}R(x,y)\sqrt{T}}
\ee
So we have now phrased the X-ray detectability as a constraint on the
gamma-ray flux, which is convenient because most of the modelling of GRB
populations is done in terms of the latter quantity. To account for the
fact that
$\Phi\sub{WFC}$ has a distribution of values rather than
a fixed value, we treat a burst at a
given $(x,y)$ with a given $P_\gamma$ as being detectable with probability
\be{eq.pdetdef}
   P\sub{det} = \frac{1}{54}\sum_{i=1}^{54}S(P_\gamma-P_{\gamma,i}\spr{min}),
\ee
where $S(x)$ is the Heaviside step function and we have defined
$P_{\gamma,i}\spr{min}$ as the minimum detectable flux for the $i$th sample
member. In other words, we just add up the fraction of model spectral
shapes in the Band set for which its X-ray flux is above threshold.  Since
we do Monte Carlo simulations to find the rate of GRB detection
(\Sect{coro}) we can then simply add up the $P\sub{det}$ values in each sky
and flux bin to get the detected rates.

In principle, the signal-to-noise of an off-centre source will be higher 
once it has been located, because the background need only be taken over the
part of the detector that is illuminated by the source. However, we will not
be aware of such sources until they are first noted in the full data stream
so this does not change the detection rate.

For the preliminary investigation in this paper, we shall use $T=20$\,s
and $f=0.5$. This means we concentrate on the long part of the bimodal
duration distribution of the bursts, which contains about 80\% of the ones
detected by BATSE. More advanced search techniques are clearly possible.
For example, one can use a number of trial values of $T$ or even trial
sky positions for the bursts to enhance the sensitivity of the search. At the
same time, this would increase the number of attempts at detection and
therefore require a higher signal-to-noise threshold to avoid spurious
signals. It is not meaningful in our view to explore these possibilities
here because the optimal strategy will depend on details of the data set
we will obtain. There will be real signals from other sources, such as
X-ray bursts and flare stars, that may well constitute a higher
contaminating rate than the expected foreground of Galactic GRB. A good
fraction of those should be discernible from GRB on spectral or temporal
grounds, but near the threshold complete subtraction may not be possible.
Also, there could be instrumental effects that very occasionally give a
spurious signal, and the importance thereof and consequences for the
search and analysis strategy are hard to predict.

\begin{figure}
   \epsfxsize=\columnwidth\epsfbox{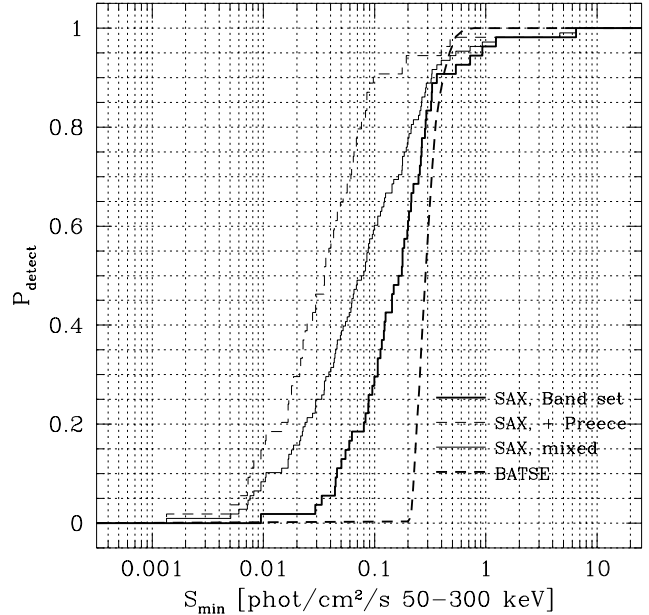}
   \caption{The detection probability of GRB at the centre
	    of the WFC field of view compared with the BATSE trigger
	    efficiency (thick dashed curve). The thick solid curve is
	    for the Band et~al. models; the thin dashed curve is with
	    each burst given an X-ray excess suggested by the data
	    from Preece et~al.\ (see text). The thin solid curve is 
	    for a mixed population of 50\% bursts with an X-ray excess
	    and 50\% without. The integration time for the WFC curves is
	    20\,s, the background rate 130\,cts\un{s}{-1}, and the
	    detection threshold is set at 5$\sigma$.
	    \label{fi:smin}
	    }
\end{figure}

In \Fig{fi:smin} we show the detection probability for bursts at the 
centre of the detector field of view as a function of $P_\gamma$. 
The thick dashed curve is for BATSE, and the thick solid one is for the WFC
using the nominal Band spectral set. Comparison of the two shows that 
above 0.3\photcms\ BATSE is more sensitive, but the curve for the WFC extends
to well below the absolute limit for BATSE. This advantage vanishes
if we look at bursts far away from the detector centre: 10\degs\ away
along the detector diagonal the curve shifts up in flux by a factor 2.4, and 
20\degs\ away that factor has grown to 12. So the effective field of view
over which the WFC can probe fainter bursts than
BATSE has a radius of about 10\degs.

There is one more spectral effect, however, that could dramatically
increase the detected rate by the WFC: in a recent paper, Preece et~al.\
(1996)\nocite{pbppm:96} find that many GRB have an excess flux above the
Band et~al. fits at energies between 5 and 10\,keV. Inspecting their
figure 5, it appears that about half of all GRB have an excess that is a
factor 10 at 5\,keV and decreases to near zero above 10\,keV. To
investigate the possible effect of this excess, we model it as a
multiplicative factor $C\sub{Preece}$ that is constant below 5\,keV and
decreases smoothly from 10 to 1 between 5 and 12\,keV. To be precise, we
multiplied the spectrum by
\begin{eqnarray}
  C\sub{Preece}(E) & = & 10                    \hspace{2cm} E<5 \nonumber \\
	& = & 1+4.5(1+\cos\frac{\pi(E-5)}{7}) \hspace{2cm} 5<E<12 \nonumber \\
		   & = & 1                     \hspace{2cm} 12<E
\end{eqnarray}
(with $E$ in keV.)
Adding this excess to each of the 54 spectra and 
recalculating the detection probabilities, we find that the entire
sensitivity curve shifts down in flux by a factor of 7 (thin dashed curve
in \Fig{fi:smin}).
Since only about half the bursts may have this excess, a more realistic
case would be an even mixture of bursts with and without excess (thin
solid curve). We stress that the fraction, $f\sub{ex}$,
of bursts with excesses and the form of the excess are quite
uncertain, so we explore the range $0.1<f\sub{ex}<0.9$ here.
But since we will have spectra of all detected bursts in the eventual 
observations we can consistently account for it in the data, since even
if no excess of GRB from M31 is found, a population of bursts with X-ray
excesses would also greatly increase the foreground rate of GRB from
our own Galaxy. Therefore we are in no danger of eventually overestimating
the constraints on halo models from WFC data.

   \section{Corona models}
   \label{coro}

For standard-candle gamma-ray bursts, the rate
density of observable bursts as a function of distance from the centre of
the corona follows directly from the observed $N(>P)$. It is usually
approximated as
\be{eq.rhoGRB}
  \rho(R) = \frac{\rho_0}{1+(R/R\sub{c})^\alpha}
\ee
and has three parameters, the central rate density $\rho_0$, the core
radius $R\sub{c}$, and the exponent $\alpha$. Since the best values of $\alpha$
are small enough that the integrated density does not converge at large $R$,
we have to add as a fourth parameter a cutoff radius of the corona, 
$R\sub{o}$, beyond which $\rho=0$. An additional model parameter is the standard-candle gamma-ray
photon emission rate $\dot{N}_\gamma$. It turns out that we only need two
of these, $R\sub{c}$ and $R\sub{o}$, because the others follow from them 
if we use known observational constraints. Moreover, the results depend only
weakly on $R\sub{o}$. We briefly indicate how a complete model is defined
once the two radii are given: First, we note that the break in the counts
slope from $-1.5$ to a smaller value occurs at 
$P\sub{break}=20\un{phot}{}\un{cm}{-2}\un{s}{-1}$. Since a burst with 
this flux is at distance $R\sub{c}$, this fixes the standard-candle value
as $\dot{N}_\gamma=4\pi R\sub{c}^2P\sub{break}$. Next we use the fact that
BATSE observes 300 bursts per year per $4\pi$ steradians above
$P\sub{comp}=1\un{phot}{}\un{cm}{-2}\un{s}{-1}$.
Because this flux is 20 times less than the break flux, it corresponds to
a distance $R\sub{c}\sqrt{20}$, and we fix $\rho_0$ by requiring the
integrated rate up to that distance to be 300/yr. $\alpha$ follows from the
fact that $N(>P_\gamma)\propto P_\gamma^{-0.7}$ at the faint end of the BATSE 
distribution. Because the asymptotic counts slope at low fluxes
for \Eq{eq.rhoGRB} is $(3-\alpha)/2$, this implies $\alpha=1.6$. This completes
the model for given $R\sub{c}$ and $R\sub{o}$.

The reasonable range of $R\sub{c}$ and $R\sub{o}$ to explore is also
limited by data. First, for $R\sub{c}\lsim30-40\,$kpc BATSE would have
detected an anisotropy due to our offset from the Galactic centre 
(Briggs et~al.\ 1996b),
and for $R\sub{c}\gsim70\,$kpc it would have seen M31. A minimum value for
the outer radius follows from the fact that BATSE sees no sign of a
truncation down
to its 50\% completeness limit of $0.28\un{phot}{}\un{cm}{-2}\un{s}{-1}$.
The distance to bursts of this flux is $8.5R\sub{c}$, i.e.\ at least
250\,kpc, so we conclude that $R\sub{o}\gsim250$\,kpc. Due to the low
sensitivity of our results to $R\sub{o}$ we fix it at half the distance 
to M31 in our calculations, noting that this is by no means required by
all models. (For neutron-star ejection models with beaming, we use 
$R\sub{o}=2$\,Mpc; see below.)
\begin{figure*}
\begin{minipage}[b]{0.75\textwidth}
   \epsfxsize=\textwidth\epsfbox{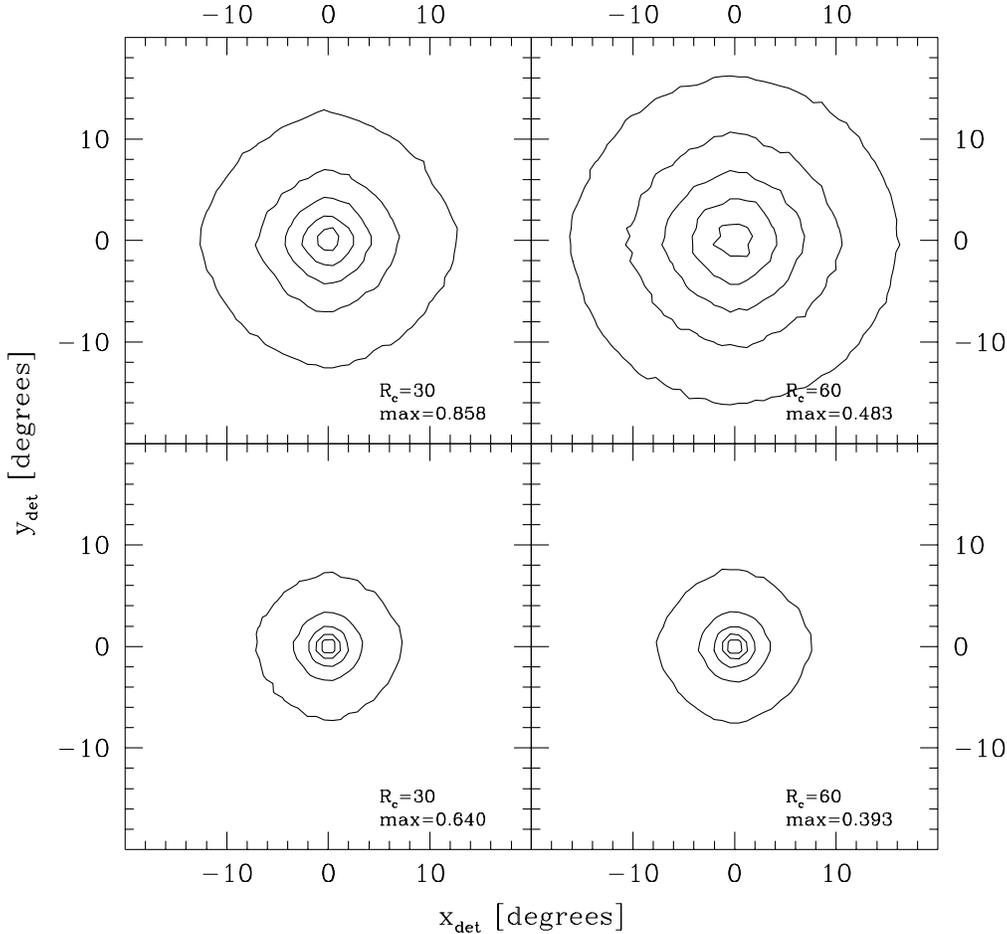}
\end{minipage}\hfill\begin{minipage}[b]{0.23\textwidth}
  \caption{Four images of the rate of detectable bursts from M31. 
	   Contours at at 10, 30, 50, 70, and 90\% of the peak rate,
	   which is indicated as `max' in each panel (in units of
	   \un{deg}{-2}\un{yr}{-1}). The x and y 
	   are offsets from the detector centre in degrees.
	   All maps are made assuming 50\% of GRB have an X-ray excess.
           \label{fi:maps}
           }
\end{minipage}
\end{figure*}

As stated, all standard-candle galactic-corona models have to satisfy this
model for the effective rate density. Since non-standard-candle models have
greater difficulty satisfying the BATSE constraints and also lead to easier
detection of M31 we shall not consider them here. However, the net rate
density is obtained in very different ways by different models, and this
has some effect on detecting M31. It does not matter whether a static halo
is used or one in which the bursters fly out of the Galaxy at high
velocity, because all models of the latter type contain a provision of
gradual or delayed turn-on of the bursting mechanism to ensure that the net
rate density becomes the same again. However, some models (e.g.\ Duncan 
et~al.\ 1993)\nocite{dlt:93}
%
%
invoke beaming of emission from fast neutron stars along their
velocity direction to avoid anisotropy. The opening angle of the beams
required in such models is of order the distance $R\sub{GC}$ from us to the
Galactic centre divided by $R\sub{c}$. This does matter, because it means
that only bursters in M31 that move roughly towards or away from us will be
seen. While this reduces the number of visible bursts, it also limits
region of the sky where they are seen to a circle of angular radius about
$R\sub{c}\sin\theta/D$ around the centre of M31, where $\theta$ is the
opening angle of the beaming cone and $D$ is the distance to M31. Since
$\theta\simeq R\sub{GC}/R\sub{c}$ the angular radius is $R\sub{GC}/D$,
independent of core radius. This area is quite small (typically a few
square degrees) and therefore the background in it is very small, which may
compensate for the lower expected rate to still give a detectable excess.

In our practical implementation, we used a Monte Carlo algorithm to create
maps of the detectable
rate of GRB per square degree per year from M31 and the Galactic
foreground given a WFC pointing direction.
The algorithm picks a burst in the Galaxy or M31 from the integrated density
distribution between $R=0$ and $R=R\sub{o}$ and random angular coordinates.
Then we compute $P_\gamma$ and the position on the sky, and from that the
detection probability for the WFC. This probability is then added to the total
in the appropriate sky location and/or flux bin and the procedure is repeated
until the rate maps have sufficiently low Monte Carlo errors.
In the case of beamed models, we also check whether the burst is shining
in our direction before counting it as detectable.  For this we make the
approximation that GRB are ejected from a galaxy at such high velocities that
they move at constant velocity in straight lines. We also assume they are all
born at the centre of the galaxy and that their emitted flux is constant for
all directions close enough to the direction of motion, and zero outside some
critical angle. This simplification allows us to check the detectability of
the burst using only the angle between our line of sight to the burst and the
direction from the burst to the centre of M31. (This approximation is quite
good, because the neutron star formation rate drops exponentially from the
centre with a small scale length of only about 4\,kpc, leading to an extra
`smearing' of the maps with an angular width of only $\theta\sub{smear}\sim
0.3^\circ$.) The opening angles (axis to edge)
of the beaming cones we used for $R\sub{c}=$30, 40, 50, 60, and 70\,kpc
are 20, 15, 12, 10, and 8 degrees, respectively, to keep $R\sub{c}\theta$
constant. Also, for these models we used outer radii of 2\,Mpc for the halos.
(They should be unlimited, but beyond that virtually
no objects will be detectable.)
\begin{figure*}
\begin{minipage}[b]{0.75\textwidth}
   \epsfxsize=\textwidth\epsfbox{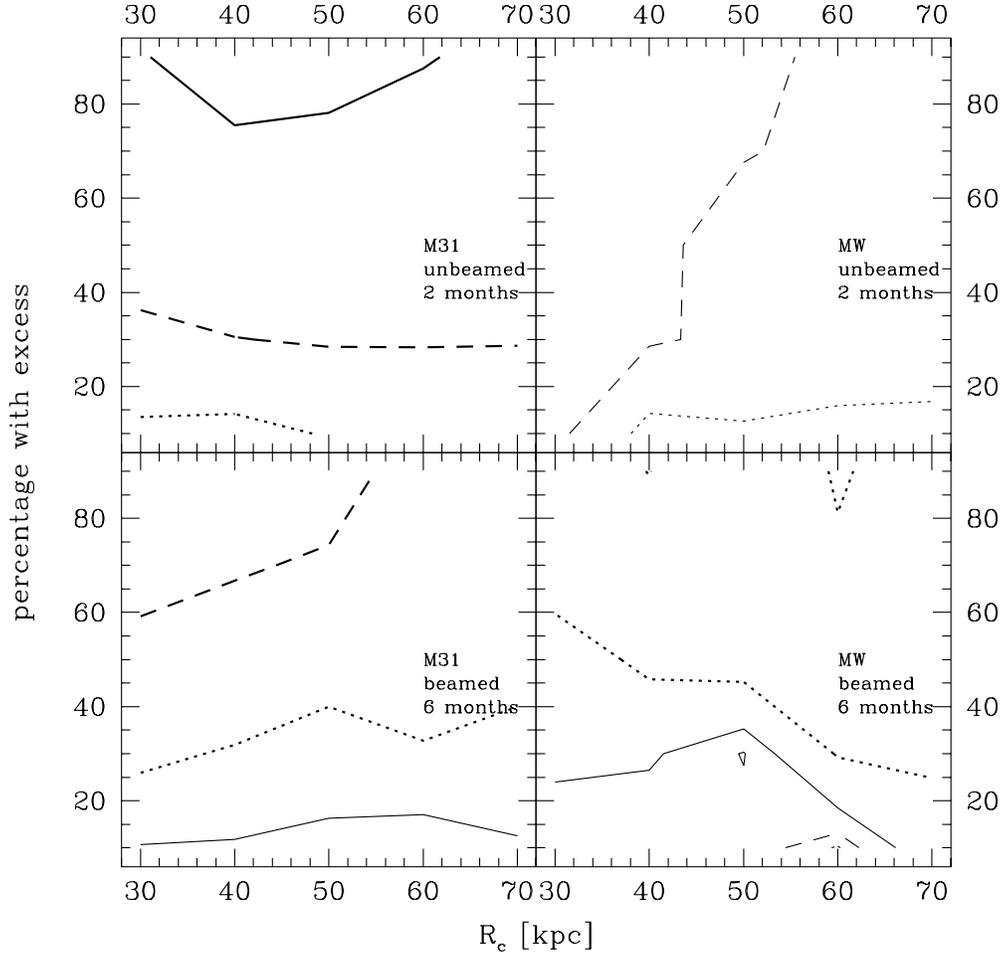}
\end{minipage}\hfill\begin{minipage}[b]{0.23\textwidth}
  \caption{Contours of the expected number of bursts in a 2-month observation
	   of unbeamed GRB (left) and a 6-month one of beamed GRB. The
	   horizontal axis gives $R\sub{c}$, the vertical one the
	   percentage of GRB with an X-ray excess. The contour values are
	   1 (thin-dotted), 2 (thin-dashed), 4 (thin-solid), 8
	   (thick-dotted), 16 (thick-dashed), and 32 (thick-solid). Left
	   is the number from M31, right that from the Galactic
	   foreground.
           \label{fi:numbers}
           }
\end{minipage}
\end{figure*}

   \section{Results}
   \label{resu}

There are three main variables on which the detectability of an excess of GRB
to M31 depends most: the core radius, the fraction of bursts, $f\sub{ex}$,
 with an X-ray
excess, and beaming. To illustrate the dependence on $R\sub{c}$ and beaming,
we show in \Fig{fi:maps} maps of the detectable rate of GRB as a function of 
position on the sky (in number per square degree per year). The top panels
are for unbeamed bursts, the bottom ones for beamed bursts. The influence
of $R\sub{c}$ is reflected by the differences between the left (30\,kpc)
and right (60\,kpc) panels. As noted above, the image becomes much smaller
for beamed bursts and the size is independent of $R\sub{c}$, whereas for 
unbeamed bursts the increase of the image size with $R\sub{c}$ is clear.
Also note the increased peak rate (labelled `max' in each panel) for
smaller core radii. This is because the central density scales as
$R\sub{c}^{-3}$, and we do see a fair fraction of the bursts at the centre
of M31. For comparison, the detectable rate from our own Milky Way is
about 0.03/deg$^2$/yr without beaming to 0.1/deg$^2$/yr with beaming (the
latter is greater due to the larger assumed outer radius for beamed models).

Because beaming so much reduces the area of sky over which GRB can be
seen, the total rate in the field of view will be much less for beamed
bursts.  This is very apparent in \Fig{fi:numbers}, in which we show the
expected number of bursts as a function of parameters. The top panels show
the expected numbers in a 2-month WFC camera observation with M31 in the
centre of the field of view for unbeamed bursts. Notice how the number
from M31 always dominates strongly. In the beamed models (bottom panels),
however, the numbers from both are comparable and a much longer
observation is needed to detect an excess towards M31.

\begin{figure*}
\begin{minipage}[b]{0.75\textwidth}
   \epsfxsize=\textwidth\epsfbox{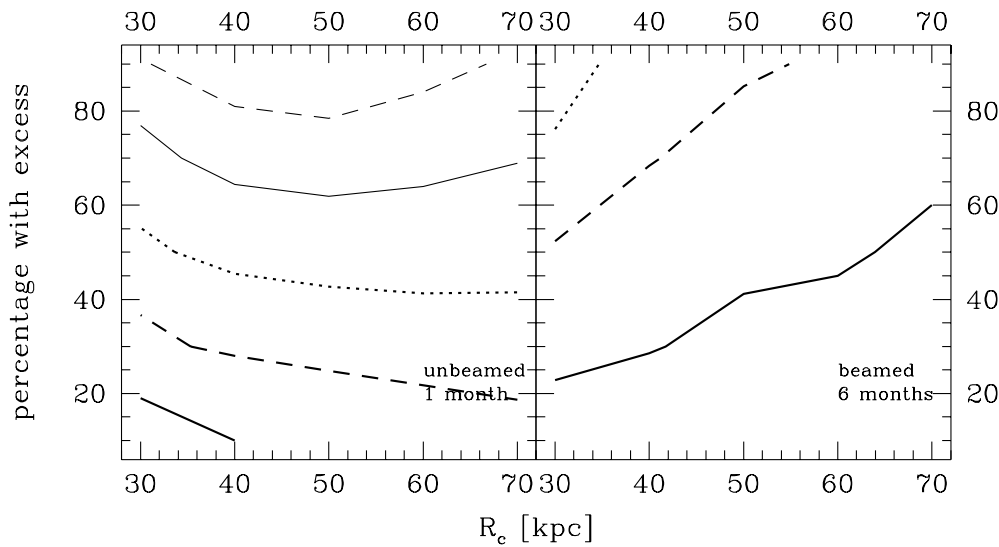}
\end{minipage}\hfill\begin{minipage}[b]{0.23\textwidth}
  \caption{Contours of the probability of making the wrong decision based
           on a 1-month observation of unbeamed GRB (top) and a 6-month one
           of beamed GRB. The horizontal axis gives $R\sub{c}$, the
           vertical one the percentage of GRB with an X-ray excess,
           $f\sub{ex}$. The contour values range from 0.001 (thin-dashed)
           to 0.1 (thick-solid) in steps of $\sqrt{10}$.
           \label{fi:perr}
           }
\end{minipage}
\end{figure*}

Now we must create a practical test of whether seeing a certain number of
GRB towards M31 constitutes evidence for or against a halo of GRB around
it. Since the foreground bursts are more spread out on the sky than those
of M31, this entails finding the optimum area of the detector to use as the
region within which we look for bursts. A good choice of boundary turns out to be
a contour on which the summed rate of M31 and Milky Way bursts is a fixed
fraction of the summed peak rate. For a given set of halo model parameters,
we then fix an observing time and a boundary. This completely specifies the
expected number of bursts, $E\sub{MW}$, if there is no halo around M31, and
the expected total number if there is, $E\sub{MW+M31}$.  The actual numbers
we would get in an observation, $N\sub{MW}$ or $N\sub{MW+M31}$, have a
Poisson distribution around the expected values.  Let us choose a
threshold value $N\sub{th}$, which defines the boundary between the
accepting the null hypothesis $H_0$ `there is no halo with these
parameters' and accepting the alternative $H_1$ `there is a halo with these
parameters'. Obviously we can make two kinds of wrong decision: reject a
halo when there is one, which happens with probability
$P(N<N\sub{th}|E\sub{MW+M31})\equiv P_1$, and accept one if there is none,
which happens with probability $P(N\ge N\sub{th}|E\sub{MW})\equiv P_2$.  The
best test will have small values of $P_1$ and $P_2$; Since we have no
preference for either type of error above the other, we shall define
$N\sub{th}$ as the value for which $P\sub{err}=\max(P_1,P_2)$ is as small
as possible. (The two can never be made equal because $N\sub{th}$ can only
assume integer values.) For a given halo model and observing time we can
then further optimise the test by varying the boundary of the detector
region inside which we accept bursts to get the lowest overall $P\sub{err}$.
In practice the optimal contour boundary lies at a rate of 5--10\% of the 
peak rate (but the minimum is shallow).

In \Fig{fi:perr} we show contours
of optimised $P\sub{err}$ values, again as a function of $R\sub{c}$ and 
$f\sub{ex}$. We can see that for the nominal value $f\sub{ex}=0.5$ we get
a quite decisive test with $P\sub{err}<0.01$ for all core radii if bursts
are not beamed. The reason that the results do not depend as much on core
radius as one might have thought is that sampling distance is not the major
issue: the X-ray brightest bursts can be seen out to a few Mpc, well beyond
M31. An increase in core radius will decrease the luminosity of bursts, but it
will also increase the central density of the halo and concentrate the 
bursts more near the detector centre. The net effect is not large.

If GRB are beamed, $P\sub{err}$ falls in the range 0.03--0.1 even for a
6-month observation, and the case is rather less convincing.  Should the
bursts be strongly beamed, our results point to a very cheap (in space
dollars) and useful satellite mission that can settle the issue: a camera
similar to the WFC, but with half the field of view and therefore only one
quarter of the background. It should be pointed to M31 for 3--6 months 
and then in some other direction for a similar period of time and would
either detect a corona of beamed sources or rule it out. Since many known
X-ray sources in M31 would be easily detected in a small fraction of the
required observing time by the same instrument, there would be considerable
benefit to such a mission for the study of variability and population analysis
of bright X-ray binaries in M31.

   \section{Discussion and conclusion} 
   \label{disc}

Previous work on constraining coronal models of GRB has many similarities
with our own calculations. Liang (1991)\nocite{liang:91} found that ROSAT
might detect some GRB in X rays assuming a now abandoned disc model for
the distribution of GRB. Li, Fenimore, \& Liang (1996)\nocite{lfl:96}
used a similar
method to our own both for beamed and unbeamed models. Their calculations
differ from ours mostly in that they use neither the spread in X-ray
luminosity nor the X-ray excesses. Their hypothetical instrument had
$S\sub{min}=0.1\photcms$. As we can see from \Fig{fi:smin}, 30--90\% of the 
bursts that the WFC can see on-axis are below this limit, so it is no
surprise that we find more optimistic prospects for detecting GRB in M31.

Harrison and Thorsett (1996)\nocite{ht:96} considered a variety of real
instruments, calculating the detectable rate in much the same way as we did
(including the spectral variability using the same set of spectra from 
Band et~al.\ 1993). They conclude that only a novel instrument
sensitive to photons in the 10--200\,keV range and with a field of view of
18$^\circ$ would be capable of detecting M31 in one year. While they did
not include the possibility of X-ray excesses, they would without doubt
have realised the potential of the SAX WFC if they had included them in their
work.

In summary, we have shown that the hitherto neglected spread in X-ray to
gamma-ray luminosity ratios of gamma-ray bursts 
substantially increases
the prospects for deciding the gamma-ray burst distance scale. The case is
further improved greatly by the recent discovery that a substantial
fraction of gamma-ray bursts have X-ray excesses (Preece et~al.\ 1996). A
one-month observation of M31 with an existing instrument, the SAX WFC,
will be decisive for establishing whether or not the Andromeda Nebula harbours
a population of bursters, unless bursters only emit radiation in fairly
narrow cones along their direction of motion. In that case, a dedicated, cheap
mission similar to the SAX WFC should resolve the issue in about one
year of observing time. Observing proposals to do
the experiment in WFC secondary (i.e.\ unguaranteed) time have been
accepted, so the gamma-ray burst distance scale
may not remain uncertain much longer.

   \section*{Acknowledgements}
   
MR thanks Steve Bell and the RGO summer school programme for the opportunity
to work in Cambridge, and the IoA for hospitality.
RAMJW thanks Arvind Parmar for pointing out the existence
of the SAX WFCs to him and Martin Rees for very interesting discussions.
This work was supported in part by grants BST 501/96 and 2~P304~016~07/94
(MR) and by PPARC (RAMJW).


\end{document}